\documentclass[aps,prl,showpacs,floatfix,twocolumn,byrevtex,superscriptaddress]{revtex4-1}
\usepackage{amsmath}
\usepackage{amssymb}
\usepackage{amstext}
\usepackage{amsopn}
\usepackage{amsfonts}
\usepackage{amsxtra}
\usepackage[english]{babel}
\usepackage{graphicx}
\usepackage{float}
\usepackage{bm}
\usepackage{multirow}
\usepackage{dcolumn}
\usepackage{color}
\usepackage{hyperref}

\newcommand{\icm}{\ensuremath{\textrm{cm}^{-1}}}

\newcommand{\KCFAF}{KCa$_{2}$Fe$_{4}$As$_{4}$F$_{2}$}
\newcommand{\KCFNAFx}{KCa$_{2}$(Fe$_{1-x}$Ni$_{x}$)$_{4}$As$_{4}$F$_{2}$}

\newcommand{\CCFAF}{CsCa$_{2}$Fe$_{4}$As$_{4}$F$_{2}$}

\newcommand{\ACFAF}{$A$Ca$_{2}$Fe$_{4}$As$_{4}$F$_{2}$}
\newcommand{\BKFAx}{Ba$_{1-x}$K$_{x}$Fe$_{2}$As$_{2}$}
\newcommand{\EF}{$E_{\text{F}}$}

\begin{document}

\title{Pseudogap and Strong Pairing Induced by Incipient and Shallow Bands in the Quasi-Two-Dimensional KCa$_{2}$Fe$_{4}$As$_{4}$F$_{2}$}
\author{Jiahao~Hao}
\thanks{These authors contributed equally to this work.}
\affiliation{National Laboratory of Solid State Microstructures and Department of Physics, Collaborative Innovation Center of Advanced Microstructures, Nanjing University, Nanjing 210093, China}
\author{Wenshan~Hong}
\thanks{These authors contributed equally to this work.}
\affiliation{Beijing National Laboratory for Condensed Matter Physics, Institute of Physics, Chinese Academy of Sciences, Beijing 100190, China}
\affiliation{School of Physical Sciences, University of Chinese Academy of Sciences, Beijing 100190, China}
\author{Xiaoxiang~Zhou}
\author{Ying~Xiang}
\author{Yaomin~Dai}
\email{ymdai@nju.edu.cn}
\affiliation{National Laboratory of Solid State Microstructures and Department of Physics, Collaborative Innovation Center of Advanced Microstructures, Nanjing University, Nanjing 210093, China}
\author{Huan~Yang}
\affiliation{National Laboratory of Solid State Microstructures and Department of Physics, Collaborative Innovation Center of Advanced Microstructures, Nanjing University, Nanjing 210093, China}
\author{Shiliang~Li}
\affiliation{Beijing National Laboratory for Condensed Matter Physics, Institute of Physics, Chinese Academy of Sciences, Beijing 100190, China}
\affiliation{School of Physical Sciences, University of Chinese Academy of Sciences, Beijing 100190, China}
\affiliation{Songshan Lake Materials Laboratory, Dongguan, Guangdong 523808, China}
\author{Huiqian~Luo}
\email{hqluo@iphy.ac.cn}
\affiliation{Beijing National Laboratory for Condensed Matter Physics, Institute of Physics, Chinese Academy of Sciences, Beijing 100190, China}
\affiliation{Songshan Lake Materials Laboratory, Dongguan, Guangdong 523808, China}
\author{Hai-Hu~Wen}
\email{hhwen@nju.edu.cn}
\affiliation{National Laboratory of Solid State Microstructures and Department of Physics, Collaborative Innovation Center of Advanced Microstructures, Nanjing University, Nanjing 210093, China}

\date{\today}
%
%

\begin{abstract}
The optical properties of KCa$_{2}$Fe$_{4}$As$_{4}$F$_{2}$ (K12442, $T_c = 33.5$~K) and KCa$_{2}$(Fe$_{0.95}$Ni$_{0.05}$)$_{4}$As$_{4}$F$_{2}$ (Ni-K12442, $T_c = 29$~K) have been examined at a large number of temperatures. For both samples, a nodeless superconducting gap is clearly observed in the optical conductivity at 5~K. The superconducting gap $\Delta \simeq 8.7$~meV ($2\Delta/k_{\text{B}}T_{c} \simeq 6.03$) in K12442, pointing towards strong-coupling Cooper pairs, but in sharp contrast, $\Delta \simeq 4.6$~meV ($2\Delta/k_{\text{B}}T_{c} \simeq 3.68$) in Ni-K12442, which agrees with the BCS weak-coupling pairing state. More intriguingly, below $T^{\ast} \simeq 75$~K, the optical conductivity of K12442 reveals a pseudogap that smoothly evolves into the superconducting gap below $T_{c}$, while no such behavior is detected in the electron-doped Ni-K12442. The comparison between the two samples hints that the pseudogap and strong-coupling Cooper pairs in K12442 may be intimately related to the shallow and incipient bands. We provide arguments supporting a preformed pairing mechanism of the pseudogap, but at the moment a magnetic scenario can not yet be excluded.
\end{abstract}



\maketitle

%
%
The recently discovered 12442-type Fe-based superconductors (FeSCs) \ACFAF\ ($A$ = K, Rb, or Cs) with a $T_{c} \simeq$ 28--33.5~K have attracted considerable attention~\cite{Wang2016JACS,Ishida2017PRB,Wang2017SCM,Wang2019PRB,Pyon2020PRM,Wang2019JPCC,Wang2019SCPMA,Hong2020PRL,Yu2019PRB,Wu2020PRB,Smidman2018PRB,Duan2021PRB,Chen2021PRL,Yi2020NJP,Chu2020CPL}. These compounds consist of double Fe$_{2}$As$_{2}$ layers separated by insulating Ca$_{2}$F$_{2}$ layers, resulting in a quasi-two-dimensional (quasi-2D) layered structure with a resistivity anisotropy of $\rho_{c}(T)/\rho_{ab}(T) \sim 10^{3}$ at low temperatures~\cite{Wang2019PRB,Pyon2020PRM}, which is significantly larger than that of FeSe ($\sim$3--4)~\cite{Vedeneev2013PRB}, LiFeAs ($\sim$1--3)~\cite{Song2010APL} and the BaFe$_{2}$As$_{2}$ family ($\sim$2--6)~\cite{Tanatar2009PRB,Tanatar2009PRB1,Nakajima2018PRB}, but comparable to that of the high-$T_{c}$ cuprates~\cite{Komiya2002PRB,Watanabe1997PRL}. Angle-resolved photoemission (ARPES) measurements on KCa$_{2}$Fe$_{4}$As$_{4}$F$_{2}$ (K12442) have observed bilayer band splitting~\cite{Wu2020PRB} analogous to that in Bi$_{2}$Sr$_{2}$CaCu$_{2}$O$_{8+\delta}$~\cite{Feng2001PRL,Chuang2001PRL}. An inelastic neutron scattering study has revealed a 2D spin resonant mode with downward dispersion in K12442~\cite{Hong2020PRL}, also resembling the behavior in cuprates.

\begin{figure}[tb]
\includegraphics[width=\columnwidth]{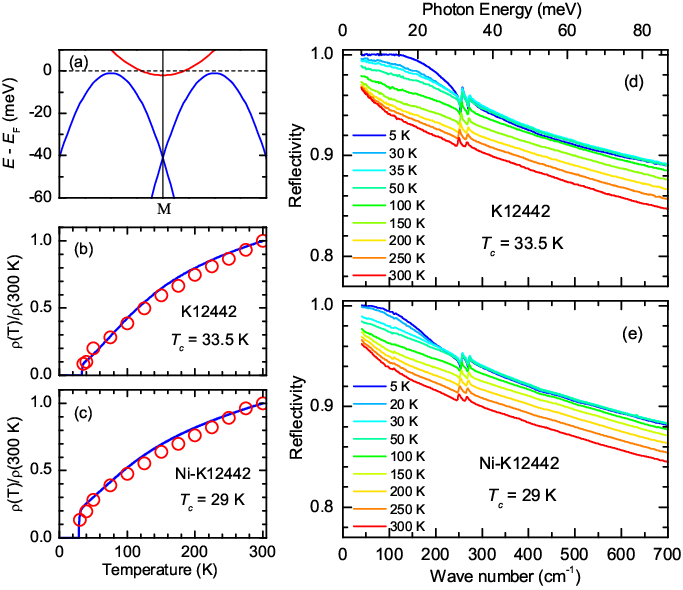}
\caption{(a) The schematic band structure of K12442 near the $M$ point of the Brillouin zone. (b) and (c) show $\rho(T)/\rho(300~\text{K})$ from transport measurements (blue solid curve) which is compared to the values extracted from our optical data (open circles) for K12442 and Ni-K12442, respectively. (d) and (e) The reflectivity of K12442 and Ni-K12442, respectively, at several representative temperatures.}
\label{PRef}
\end{figure}

More interestingly, as schematically shown in Fig.~\ref{PRef}(a), near the $M$ point of the Brillouin zone, K12442 has a very shallow electron band, whose bottom barely touches $E_{\text{F}}$, and four incipient hole bands with their tops being very close to but not crossing $E_{\text{F}}$~\cite{Wu2020PRB}. This kind of band topology places the system near a Lifshitz transition, which may induce an $s + is$ pairing state with broken time-reversal symmetry~\cite{Maiti2013PRB,Boker2017PRB,Grinenko2020NP,Grinenko2021NP}. In addition, the shallow and incipient bands lead to very large $\Delta/E_{\text{F}}$ ratios. Here, $\Delta$ and $E_{\text{F}}$ correspond to the superconducting (SC) gap and the Fermi energy, respectively. A large $\Delta/E_{\text{F}}$ ratio ($\Delta/E_{\text{F}} \sim 1$) may drive the system into the crossover regime between the weak-coupling Bardeen-Cooper-Schrieffer (BCS) and the strong-coupling Bose-Einstein condensation (BEC) limits~\cite{Randeria2014ARCMP,Nakagawa2021Science,Chubukov2016PRB,Hashimoto2020SA,Lubashevsky2012NP,Okazaki2014SR,Rinott2017SA,Kasahara2016NC}. One of the most intriguing phenomena in the BCS-BEC crossover regime is the opening of a pseudogap due to preformed Cooper pairs between the pairing temperature $T^{\ast}$ and the critical temperature $T_{c}$ at which the pairs condense into a phase-coherent quantum state~\cite{Randeria2014ARCMP,Nakagawa2021Science,Chubukov2016PRB,Hashimoto2020SA,Tajima2019PRB,Tajima2020PRB}. Although pseudogap behavior has been widely reported in FeSCs~\cite{Dai2012PRB,Mertelj2009PRL,Shi2018JPSJ,Shimojima2014PRB,Yang2016PRB,Charnukha2018PRL,Charnukha2014JPCM,Kordyuk2015LTP}, its relation to the shallow or incipient band is still unclear.

Elucidating the role of the shallow and incipient bands in K12442 may shed new light on the pairing mechanism in FeSCs. This can be achieved by comparing the spectroscopic properties of K12442 and the electron-doped compound, in which $E_{\text{F}}$ is shifted away from the bottom of the shallow electron band and the tops of the incipient hole bands due to electron doping. Here, we report on a detailed optical study of K12442 ($T_c = 33.5$~K) and the electron-doped Ni-K12442 ($T_c = 29$~K). In the SC state, the optical conductivity of both samples reveals a nodeless SC gap. While the SC gap $\Delta \simeq 8.7$~meV ($2\Delta/k_{\text{B}}T_{c} \simeq 6.03$) in K12442, suggesting strong-coupling Cooper pairs, the electron-doped Ni-K12442 exhibits $\Delta \simeq 4.6$~meV ($2\Delta/k_{\text{B}}T_{c} \simeq 3.68$), falling into the BCS weak-coupling pairing regime. More interestingly, a pseudogap opens below $T^{\ast} \simeq 75$~K and evolves into the SC gap below $T_c$ in K12442, whereas such pseudogap behavior is absent in Ni-K12442. The comparison between the two samples indicates that the shallow and incipient bands may play an important role in the formation of the pseudogap and the strong-coupling SC gap in the quasi-2D K12442. While we argue that precursor superconductivity may be the origin of the pseudogap, we cannot, at the moment, rule out a magnetic mechanism.

%
%

Single crystalline KCa$_{2}$Fe$_{4}$As$_{4}$F$_{2}$ (K12442) with $T_c = 33.5$~K and KCa$_{2}$(Fe$_{0.95}$Ni$_{0.05}$)$_{4}$As$_{4}$F$_{2}$ (Ni-K12442) with $T_c = 29$~K were synthesized using the self-flux method~\cite{Wang2019JPCC,Wang2019SCPMA,Hong2020PRL}. The sample characterization and details about the experimental method can be found in the Supplementary Material~\cite{SuppMat} and Refs.~\cite{Homes1993,Dressel2002}. Figure~\ref{PRef}(d) displays the reflectivity $R(\omega)$ of K12442 in the far-infrared range at 9 selected temperatures above and below $T_{c}$. The normal-state $R(\omega)$ has a high value in this range and rises with decreasing $T$, which is the optical characteristic of metallic materials. Below $T_{c}$, an upturn emerges in the low-frequency $R(\omega)$, indicating the opening of an SC gap. $R(\omega)$ at 5~K exhibits a flat response at unity below $\sim$130~\icm, suggesting that the SC gap does not have nodes~\cite{Degiorgi1994PRB,Li2008PRL,Dai2013EPL,Xu2019PRB,Dai2016PRB}. Similar behavior is observed in $R(\omega)$ of Ni-K12442, as shown in Fig.~\ref{PRef}(e), but the upturn and the flat response in $R(\omega)$ at 5~K shift to lower energy, implying a smaller SC gap in Ni-K12442.

%
%

\begin{figure}[tb]
\includegraphics[width=\columnwidth]{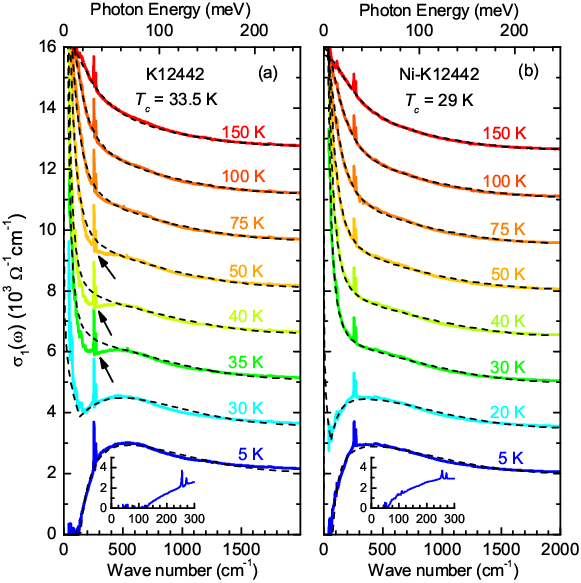}
\caption{$\sigma_{1}(\omega)$ for K12442 (a) and Ni-K12442 (b) at several selected temperatures above and below $T_{c}$. The spectra at different temperatures are shifted by 1500~$\Omega^{-1}$cm$^{-1}$ to avoid overlap. The dashed lines are fitting results. The inset of each panel shows an enlarged view of $\sigma_{1}(\omega)$ at 5~K in the low-frequency range.}
\label{PS1}
\end{figure}
Figure~\ref{PS1}(a) depicts the real part of the optical conductivity $\sigma_{1}(\omega)$ of K12442 at a variety of temperatures below and above $T_{c}$. In the normal state, e.g. $T = 150$~K, $\sigma_{1}(\omega)$ is characterized by a pronounced Drude peak centered at zero frequency, which is the optical fingerprint of a metal. The width of the Drude peak at half maximum represents the quasiparticle scattering rate. Upon cooling, the Drude peak narrows, indicating a reduction of the quasiparticle scattering rate. In the SC state, for example $T = 5$~K, the low-frequency Drude response in $\sigma_{1}(\omega)$ is completely suppressed due to the opening of the SC gap. $\sigma_{1}(\omega)$ drops to zero below $\sim$130~\icm, as shown in the inset of Fig.~\ref{PS1}(a). This is a clear signature of a nodeless SC gap~\cite{Degiorgi1994PRB,Li2008PRL,Dai2013EPL,Xu2019PRB,Dai2016PRB}. Furthermore, it is noteworthy that a gap feature, indicated by the arrows in Fig.~\ref{PS1}(a), develops in $\sigma_{1}(\omega)$ below $T^{\ast} \simeq 75$~K. Such a normal-state gap (pseudogap) intensifies as $T$ is lowered, and finally evolves into the SC gap below $T_{c}$. By comparison, the normal-state $\sigma_{1}(\omega)$ for Ni-K12442 [Fig.~\ref{PS1}(b)] exhibits standard Drude behavior down to 30~K without an evident pseudogap feature. Moreover, as displayed in the inset of Fig.~\ref{PS1}(b), $\sigma_{1}(\omega)$ of Ni-K12442 at 5~K vanishes below $\sim$70~\icm, suggesting that the SC gap in Ni-K12442 is also nodeless, but much smaller than that in K12442. Here, we should mention that while ARPES~\cite{Wu2020PRB}, optical spectroscopy~\cite{Xu2019PRB}, scanning tunneling microscopy (STM)~\cite{Duan2021PRB} and heat transport~\cite{Huang2019PRB} studies have revealed multiple nodeless SC gaps in the 12442 compounds, nodes in the SC gaps have been suggested by muon-spin rotation ($\mu$SR)~\cite{Kirschner2018PRB,Smidman2018PRB} and specific heat~\cite{Wang2020SCPMA} measurements. These inconsistencies may be interpreted by the existence of accidental nodes in some $k_{z}$ planes in the Brillouin zone~\cite{Zhang2012NP}.

\begin{figure}[tb]
\includegraphics[width=\columnwidth]{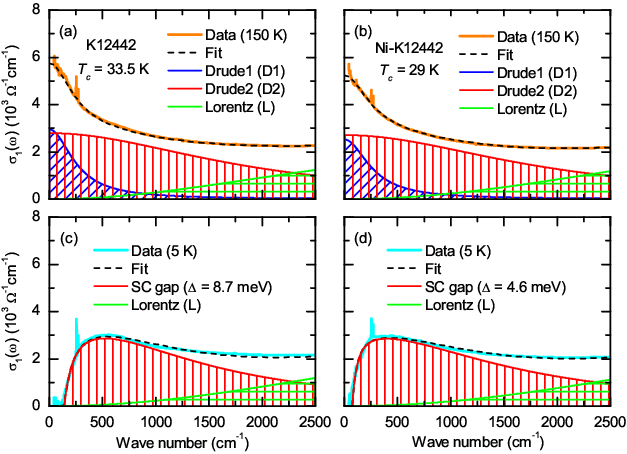}
\caption{(a) The orange curve is $\sigma_{1}(\omega)$ of K12442 at 150~K. The dashed line represents the fit which is decomposed into a narrow Drude D1 (blue hatched area), a broad Drude D2 (red hatched area), and a Lorentz component L (green hatched area). (b) shows the same curves for Ni-K12442. (c) and (d) The modeling results at $T = 5$~K for K12442 and Ni-K12442, respectively. Both consist of an SC gap (red hatched area) and the same L as that in the normal-state fit.}
\label{PFit}
\end{figure}
We fit the normal-state $\sigma_{1}(\omega)$ to the Drude-Lorentz model (Supplementary Material)~\cite{SuppMat}. The orange curve in Fig.~\ref{PFit}(a) is $\sigma_{1}(\omega)$ of K12442 at 150~K, and the black dashed line represents the fit, which consists of a narrow Drude D1 (blue hatched area), a broad Drude D2 (red hatched area), and a Lorentz component L (green hatched area). A similar fit is obtained for $\sigma_{1}(\omega)$ of Ni-K12442 at 150~K as shown in Fig.~\ref{PFit}(b). The fitting parameters for both compounds are listed in Tab.~S1 in the Supplementary Material. Such a two-Drude model has been widely used to describe the low-frequency $\sigma_{1}(\omega)$ of FeSCs in the normal state~\cite{Wu2010PRB,Dai2016PRB,Dai2013PRL,Nakajima2010PRB,Xu2019PRB}. In the SC state, $\sigma_{1}(\omega)$ of FeSCs is usually modeled by replacing the two Drude components with two SC gaps~\cite{Maksimov2011PRB,Dai2013EPL,Homes2015PRB} given by the Mattis-Bardeen formalism (Supplementary Material)~\cite{SuppMat,Zimmermann1991PC}. However, as depicted in Fig.~\ref{PFit}(c), $\sigma_{1}(\omega)$ of K12442 at 5~K (cyan solid curve) can be well described by the superposition (black dashed line) of a single SC gap (red hatched area) formed in D2 and the same Lorentz term from the normal state (green hatched area); D1 vanishes in the SC state. This behavior, which has been discussed in \CCFAF~\cite{Xu2019PRB} and LiFeAs~\cite{Dai2016PRB}, signifies the coexistence of clean- and dirty-limit superconductivity in K12442, and D1 corresponds to the clean-limit SC bands. The gap values in the clean-limit bands can not be determined from $\sigma_{1}(\omega)$~\cite{Kamaras1990PRL}, because upon the SC condensate, clean-limit bands disappear from the finite-frequency $\sigma_{1}(\omega)$, leaving no observable feature at the energy of the SC gap~\cite{Dai2016PRB,Xu2019PRB,Kamaras1990PRL}. The SC gap in D2 ($\Delta \simeq 8.7$~meV) leads to a ratio of $2\Delta/k_{\text{B}}T_{c} \simeq 6.03$ that is much larger than the weak-coupling BCS value 3.52, pointing to strong-coupling Cooper pairs. Since the value of the SC gap in D2 is close to the large gap observed by other probes~\cite{Wu2020PRB,Duan2021PRB,Smidman2018PRB}, the small gaps are likely to be hidden in the clean-limit bands (D1). Figure~\ref{PFit}(d) shows the fit of $\sigma_{1}(\omega)$ at 5~K for Ni-K12442, which is essentially the same as that for K12442. Interestingly, the SC gap value $\Delta \simeq 4.6$~meV in Ni-K12442 yields a ratio of $2\Delta/k_{\text{B}}T_{c} \simeq 3.68$ close to the BCS value. This suggests that doping electrons into K12442 significantly reduces the pairing strength in some bands.

\begin{figure}[tb]
\includegraphics[width=\columnwidth]{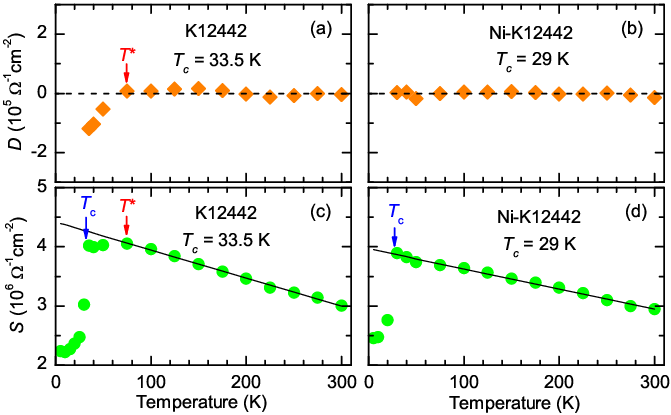}
\caption{(a)--(b) $T$ dependence of the difference between $\sigma_{1}(\omega)$ and the Drude-Lorentz fit. $T^{\ast}$ denotes the pseudogap opening temperature. (c)--(d) $T$ dependence of the spectral weight $S$. The black lines are linear guides to the eyes.}
\label{PTD}
\end{figure}
We next examine the $T$ dependence of $\sigma_{1}(\omega)$ for both materials. We apply the Drude-Lorentz analysis to $\sigma_{1}(\omega)$ for all temperatures above $T_{c}$. The fitting results at several representative temperatures are traced out as black dashed lines in Figs.~\ref{PS1}(a) and \ref{PS1}(b) for K12442 and Ni-K12442, respectively. The inverse of the zero-frequency value of the fit $1/\sigma_{1}^{\text{Fit}}(\omega = 0)$ [open circles in Figs.~\ref{PRef}(b) and \ref{PRef}(c)] agrees with transport measurements for both samples. While the two-Drude model describes $\sigma_{1}(\omega)$ of K12442 quite well above 75~K, below this temperature, such as 50, 40, and 35~K, a striking deviation between $\sigma_{1}(\omega)$ and the fit occurs due to the development of the pseudogap as indicated by the black arrows in Fig.~\ref{PS1}(a). In contrast, such a pseudogap is absent in Ni-K12442 [Fig.~\ref{PS1}(b)], and $\sigma_{1}(\omega)$ can be well described by the two-Drude model at all temperatures above $T_{c}$. Figures~\ref{PTD}(a) and \ref{PTD}(b) display the $T$ dependence of the difference $D$ between $\sigma_{1}(\omega)$ and the fit $\sigma_{1}^{\text{Fit}}(\omega)$, defined as
%
%
\begin{equation}
D = \int_{\omega_a}^{\omega_b}[\sigma_{1}(\omega)-\sigma_{1}^{\text{Fit}}(\omega)]d\omega
\label{Dev},
\end{equation}
where $\omega_{a}=40$~\icm\ and $\omega_{b}=1000$~\icm\ represent the lower and upper cut-off frequencies, respectively. For Ni-K12442 [Fig.~\ref{PTD}(b)], $D \simeq 0$ at all temperatures, indicating a good agreement between $\sigma_{1}(\omega)$ and $\sigma_{1}^{\text{Fit}}(\omega)$, whereas a decrease in $D$ occurs below $T^{\ast} \simeq 75$~K for K12442 [Fig.~\ref{PTD}(a)], signaling the opening of the pseudogap.

To further attest to the formation of the pseudogap above $T_c$ in K12442, it is informative to track the $T$ dependence of the spectral weight
%
%
\begin{equation}
S = \int_{0^+}^{\omega_c}\sigma_{1}(\omega)d\omega
\label{SW},
\end{equation}
where $\omega_{c}$ is a cut-off frequency; $0^{+}$ means that the superfluid weight, represented by a zero-frequency $\delta$ function in $\sigma_{1}(\omega)$, is not included. Figure~\ref{PTD}(c) depicts the $T$ dependence of $S$ with $\omega_{c} = 1000$~\icm\ for K12442. As $T$ is lowered from 300~K, $S$ increases continuously, following the black solid line (a linear guide to the eyes). This behavior is a direct consequence of the narrowing of the Drude response. Below $T_{c}$, a sharp drop of $S$ sets in, which is a clear signature of the SC gap opening~\cite{Dai2013EPL}. Moreover, a noteworthy phenomenon is that $S$ deviates from the high-temperature trend and starts to decrease below $T^{\ast} \simeq 75$~K, suggesting that a pseudogap starts to open below $T^{\ast}$. For Ni-K12442 [Fig.~\ref{PTD}(d)], while both the continuous increase of $S$ upon cooling in the normal state and the suppression of $S$ below $T_{c}$ are clearly observed, the signature of the pseudogap is absent.

%

\begin{figure}[tb]
\includegraphics[width=\columnwidth]{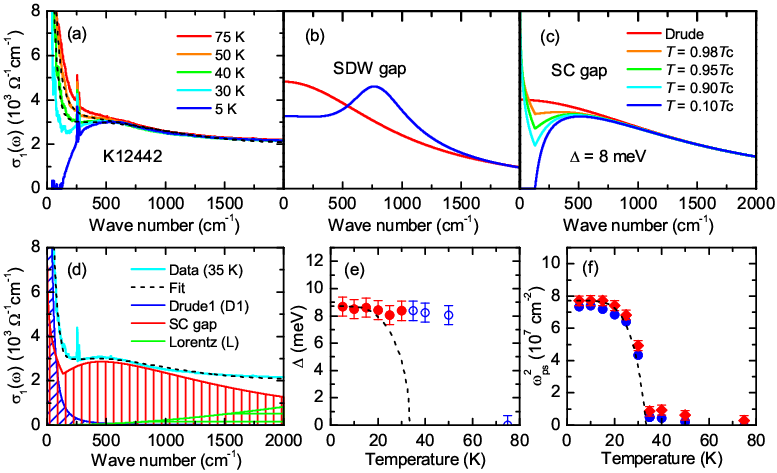}
\caption{(a) $\sigma_1(\omega)$ of K12442 at different temperatures showing the evolution of the pseudogap into the SC gap. The dashed lines are fits using an SC gap. (b) The optical response of an SDW gap. (c) The evolution of an SC gap ($\Delta$ = 8~meV) with temperature. (d) The cyan curve is $\sigma_1(\omega)$ of K12442 at 35~K. The dashed line denotes the fit which is obtained by replacing D2 in the Drude-Lorentz model with a Mattis-Bardeen term with $\Delta = 8.4$~meV and $T/T_{c} = 0.95$. (e) The values of the SC gap (red solid circles) and the pseudogap (blue open circles) as a function of temperature. The dashed line represents the BCS mean-field behavior. (f) $\omega^{2}_{\text{ps}}$ determined from $\sigma_{2}(\omega)$ (red solid diamonds) and from the missing area in $\sigma_{1}(\omega)$ (blue solid circles). The dashed line is a BCS calculation assuming an $s$-wave gap with $\Delta = 8.7$~meV.}
\label{PMBTD}
\end{figure}
Having established the existence of the pseudogap below $T^{\ast}$ in K12442, we next discuss its origin. Previous optical studies have shown that disorder causes carrier localization, creating a similar feature in the low-frequency $\sigma_{1}(\omega)$~\cite{Gaymann1995PRB,Basov1994PRB,Basov1998PRL,Kobayashi2016PRB}. In the \KCFNAFx\ system, since Ni doping introduces disorder into the FeAs layers, the Ni-doped compound has a higher degree of disorder than the stoichiometric K12442. While the pseudogap is observed in K12442, it is absent in the more disordered Ni-K12442. This rules out the disorder effect as a possible origin of the pseudogap in K12442. The parent and underdoped compounds of FeSCs often exhibit a spin-density-wave (SDW) phase or SDW fluctuations, which also open a gap in $\sigma_{1}(\omega)$~\cite{Hu2008PRL,Nakajima2010PRB,Xu2016PRB,Dai2016PRB122}. K12442, with the Fe valence being +2.25, is intrinsically self-doped to a level of 0.25 holes/Fe~\cite{Wang2016JACS,Ishida2017PRB}. This doping level corresponds to the slightly over-doped \BKFAx\ with $x = 0.5$, where no SDW order or fluctuations are expected~\cite{Xu2016PRB}. In addition, no SDW phase has been experimentally detected in K12442 or electron-doped compounds~\cite{Ishida2017PRB}. Furthermore, an SDW gap, as schematically illustrated in Fig.~\ref{PMBTD}(b), depletes the low-frequency $\sigma_{1}(\omega)$ and transfers the spectral weight to a higher frequency range~\cite{Hu2008PRL,Nakajima2010PRB,Xu2016PRB,Dai2016PRB122}. In K12442, as shown in Fig.~\ref{PMBTD}(a), the pseudogap suppresses the low-frequency $\sigma_{1}(\omega)$, but no spectral weight is transferred to the higher frequency range. These facts suggest that the pseudogap in K12442 is unlikely to be associated with an SDW order or SDW fluctuations.

Since the pseudogap smoothly evolves into the SC gap below $T_{c}$ [Fig.~\ref{PMBTD}(a)] and its evolution with $T$ mimics the $T$ dependence of the SC gap as illustrated in Fig.~\ref{PMBTD}(c), it is most likely to be a precursor of the SC gap. Driven by this possibility, we replace D2 in the Drude-Lorentz model with an SC gap to describe $\sigma_{1}(\omega)$ of K12442 between $T^{\ast} \simeq 75$~K and $T_c = 33.5$~K. As shown in Fig.~\ref{PMBTD}(d), the superposition (dashed line) of a Mattis-Bardeen term with $\Delta = 8.4$~meV and $T/T_{c} = 0.95$ (red hatched area), D1 (blue hatched area) and L (green hatched area) describes $\sigma_{1}(\omega)$ of K12442 at 35~K quite well. The same approach also reproduces $\sigma_{1}(\omega)$ at 40~K ($\Delta = 8.2$~meV, $T/T_{c} = 0.96$) and 50~K ($\Delta = 8.0$~meV, $T/T_{c} = 0.97$), as shown by the dashed lines in Fig.~\ref{PMBTD}(a). Figure~\ref{PMBTD}(e) plots the $T$ dependence of the SC gap (red solid circles) and the pseudogap (blue open circles) extracted from the Mattis-Bardeen fit. The SC gap barely changes with $T$, deviating from the BCS mean-field theory (dashed line) in the vicinity of $T_{c}$, which is consistent with ARPES measurements~\cite{Wu2020PRB} and resembles the behavior of the large gap in \BKFAx~\cite{Ding2008EPL,Dai2013EPL}. Above $T_c$, the SC gap evolves into the pseudogap which has a similar value and disappears at $T^{\ast}\simeq75$~K. The fact that the pseudogap feature in $\sigma_{1}(\omega)$ can be described by the Mattis-Bardeen profile not only lends credence to the precursor scenario, but also implies the existence of superfluid weight below $T^{\ast}$. From the viewpoint of the sum rule~\cite{Tinkham1959PRL,Ferrell1958PR}, the presence of superfluid weight which lies at zero frequency is compatible with the loss of finite-frequency spectral weight induced by the opening of the pseudogap as observed in Fig.~\ref{PTD}(c). The superfluid plasma frequency $\omega_{\text{ps}}$ ($\omega^2_{\text{ps}}=Z_{0}N_{s}/\pi^2$ where $N_{s}$ is the superfluid weight) can be determined from the imaginary part of the optical conductivity $\sigma_{2}(\omega)$ or the missing area in $\sigma_{1}(\omega)$ due to the formation of SC condensate (see Supplementary Material~\cite{SuppMat} and Refs.~\cite{Tinkham1959PRL,Ferrell1958PR,Dordevic2002PRB,Zimmers2004PRB,Uykur2014PRL,Lee2017JPSJ}). Figure~\ref{PMBTD}(f) displays the $T$ dependence of $\omega^2_{\text{ps}}$ determined from $\sigma_{2}(\omega)$ (red diamonds) and from the missing area in $\sigma_{1}(\omega)$ (blue circles), which in general agrees with the calculation (dashed line) assuming an $s$-wave gap with $\Delta$ = 8.7~meV. Interestingly, a non-zero $\omega^2_{\text{ps}}$ is obtained between $T_{c}$ and $T^{\ast}$, indicating the emergence of precursor superconductivity below $T^{\ast}$. Such a precursor superconducting state has been observed in cuprates far above $T_{c}$ but below the pseudogap temperature~\cite{Dubroka2011PRL,Uykur2014PRL,Lee2017JPSJ}. Furthermore, a recent pump-probe study reported a pseudogap below $T^{\ast} \simeq 50$~K associated with a precursor of superconductivity~\cite{Zhang2022SCPMA}.

Theoretical calculations~\cite{Chubukov2016PRB} have shown that in the 2D case, when $E_{\text{F}}$ is small (BCS-BEC crossover), the system displays pseudogap behavior due to preformed pairs above $T_{c}$. Given that K12442 is quasi-2D, and has shallow and incipient bands (very small $E_{\text{F}}$), the pseudogap probably originates from preformed Cooper pairs related to the BCS-BEC crossover~\cite{Randeria2014ARCMP,Chubukov2016PRB,Hashimoto2020SA,Lubashevsky2012NP,Okazaki2014SR,Rinott2017SA,Kasahara2016NC}. This scenario seems compatible with the doping dependence of the optical response as well. The substitution of Ni for Fe introduces electrons to the system and shifts $E_{\text{F}}$ up, resulting in a decrease of $\Delta/E_{\text{F}}$, which moves the system towards the BCS direction, accounting for the absence of the pseudogap in Ni-K12442. Moreover, our optical data have shown that doping electrons into K12442 reduces the pairing strength in some bands. This result can also be understood in the framework that K12442 is in the BCS-BEC crossover regime which is characterized by strong pairing, while Ni doping drives the system towards the BCS direction, where weak-coupling Cooper pairs dominate. A recent specific heat study places K12442 in the BCS-BEC crossover regime based on the fact that the onset point of the SC transition is almost unchanged under a magnetic field as high as 9~T~\cite{Wang2020SCPMA}.

Nevertheless, we would like to remark that challenges to the BCS-BEC crossover scenario also exist. For example, in a multiband system, whether $\Delta/E_{\text{F}}$ can be defined individually for each band is unclear, and evidence of strong SC fluctuations expected for the BCS-BEC crossover is still elusive. While recent Nernst, Hall effect, and NMR studies on \CCFAF\ have revealed anomalous behavior below $\sim$90~K which may be related to fluctuating SC~\cite{Li2022PRB}, signatures of strong SC fluctuations on such a high temperature scale are absent from specific heat and transport measurements~\cite{Wang2019PRB,Wang2019SCPMA,Wang2020SCPMA,Pyon2020PRM}. There are similar controversies over FeSe, a well-known BCS-BEC crossover superconductor candidate~\cite{Kasahara2016NC,Shi2018JPSJ,Yang2017PRB,Takahashi2019PRB,Hanaguri2019PRL}. On the one hand, giant SC fluctuations and a pseudogap have been detected well above $T_{c}$ in FeSe by multiple techniques including transport, susceptibility, magnetic torque, NMR, Hall effect, Seebeck, and Nernst effect~\cite{Kasahara2016NC,Shi2018JPSJ}. On the other hand, BCS-like behavior is found in other transport, magnetization, specific heat, and Nernst measurements~\cite{Yang2017PRB}; a recent magnetic torque measurement on FeSe points to the absence of fluctuating diamagnetism~\cite{Takahashi2019PRB}; a scanning tunneling microscopy (STM) study of FeSe fails to detect a pseudogap~\cite{Hanaguri2019PRL}, at variance with the BCS-BEC crossover. Hence, whether K12442 is a BCS-BEC crossover superconductor calls for further investigations.

Strong pairing and pseudogap behavior are observed in K12442 which has shallow and incipient bands, whereas in Ni-K12442, with the shallow and incipient bands being eliminated by electron doping, our optical study reveals BCS-like weak-coupling Cooper pairs without a pseudogap. These results imply that the pseudogap and strong pairing in K12442 are closely related to the shallow and incipient bands. A pseudogap seems to also exist in $\sigma_{1}(\omega)$ of \CCFAF~\cite{Xu2019PRB}, as evidenced by the discrepancy between the measured $\sigma_{1}(\omega)$ at 35~K and the Drude-Lorentz fit near 200~\icm, but it is much weaker than that in \KCFAF. Although the calculated band structures of \ACFAF\ ($A$ = K, Rb, or Cs) are similar~\cite{Wang2016EPL,Singh2018AIPCPCs,Singh2018AIPCPKRb}, in real materials it is natural that the difference in the alkali atom may lead to a slight difference in \EF. If the shallow and incipient bands play an important role in the formation of the pseudogap, a relatively small difference in \EF\ may have a noticeable impact on the behavior of the pseudogap in \ACFAF. A similar pseudogap feature is visible in $\sigma_{1}(\omega)$ of the optimally doped Ba$_{1-x}$K$_{x}$Fe$_{2}$As$_{2}$ reported by Refs.~\cite{Charnukha2011NC,Charnukha2011PRB}, but absent in the data measured by other groups~\cite{Li2008PRL,Dai2013EPL,Dai2013PRL,Mallett2017PRB}. Such a discrepancy may also arise from the difference in \EF.

In FeSCs, it has been generally accepted that superconductivity stems from spin-fluctuation mediated interband pairing interactions, which give rise to an $s_{\pm}$ gap symmetry~\cite{Mazin2008PRL,Kuroki2008PRL}. The SC gaps in different bands follow $\Delta_{1}/\Delta_{2} = -\sqrt{N_{2}/N_{1}}$~\cite{Hirschfeld2011RPP}, where $N_{i}$ is the total density of states contributed by the $i$th band. This equation indicates that a shallow or incipient band exhibits a large SC gap, but can only make a small contribution to superfluidity. Furthermore, previous work on FeSCs has revealed that shallow and incipient bands usually host strong pairing~\cite{Lubashevsky2012NP,Miao2015NC,Shimojima2017SA,Chubukov2016PRB,Chen2015PRB}. Considering these facts, we propose that in K12442, the interband pairing interaction involving the shallow and incipient bands induces strong pairing and a large gap near $M$ with very small superfluid weight, leading to a pseudogap that behaves like a precursor of an SC gap. In Ni-K12442, as the shallow and incipient bands near $M$ are eliminated, the interband pairing interaction between the hole pockets at $\Gamma$ and the electron pocket at $M$ induces a moderate SC gap with finite superfluid weight. In this case, no pseudogap is expected. Finally, we would like to point out that while our results are consistent with the precursor superconductivity scenario for the pseudogap, we cannot completely exclude a magnetic mechanism, such as SDW fluctuations.

%
%
To summarize, we investigated the optical properties of K12442 ($T_c = 33.5$~K) and Ni-K12442 ($T_c = 29$~K) at numerous temperatures. In both compounds, a nodeless SC gap is clearly observed in $\sigma_{1}(\omega)$ at 5~K. The SC gap $\Delta \simeq 8.7$~meV ($2\Delta/k_{\text{B}}T_{c} \simeq 6.03$) in K12442, indicating strong pairing, but in sharp contrast, $\Delta \simeq 4.6$~meV ($2\Delta/k_{\text{B}}T_{c} \simeq 3.68$) in Ni-K12442, consistent with the BCS weak-coupling pairing state. More interestingly, below $T^{\ast} \simeq 75$~K, a pseudogap develops in $\sigma_{1}(\omega)$ of K12442, and smoothly evolves into the SC gap below $T_{c}$, while no pseudogap behavior is detected in Ni-K12442. The pseudogap and strong-coupling SC gap in K12442 may be intimately related to the shallow and incipient bands. We argue that the pseudogap may be associated with preformed Cooper pairs, but at the moment a magnetic mechanism can not yet be ruled out.

%
%

\begin{acknowledgments}
We thank I.~Eremin, I.~I.~Mazin, Q.~H.~Wang, H.~Miao, B.~Xu, K.~Y.~Gao, J.~B.~Qi, and P.~Zhang for helpful discussions, Q.~Li and C.~P.~He for assistance with EDS analysis. We acknowledge financial support from the National Key Research and Development Program of China (Grants No. 2016YFA0300400, No. 2015CB921202, and No. 2018YFA0704200), the National Natural Science Foundation of China (Grants No. 11874206, No. 12174180, No. 12061131001, No. 11822411 and No. 11961160699), the Jiangsu Shuangchuang program, and the Strategic Priority Research Program (B) of the Chinese Academy of Sciences (CAS) (Grant No. XDB25000000 and No. XDB07020300). H. L. is grateful for the support from the Youth Innovation Promotion Association of CAS (Grant No. Y202001) and the Beijing Natural Science Foundation (Grant No. JQ19002).
\end{acknowledgments}

%

\end{document}